\documentclass[aps,prb,twocolumn]{revtex4-1}

\usepackage{graphicx}
\usepackage{dcolumn}
\usepackage{bm}
\usepackage{booktabs,array,mathptmx,float,tabularx,lipsum,amsmath,multirow}
\usepackage{siunitx,xcolor}
\usepackage{color}
\usepackage{booktabs}
\usepackage{epstopdf}
\usepackage{ulem}
\usepackage{verbatim}
\usepackage{subfigure}

\begin{document}


\title{An \textit{ab initio} study of magnetic structure transitions of FePS$_3$ under high pressure}


\author{Yijie Zeng}
\affiliation{College of Science, Hangzhou Dianzi University, Hangzhou 310018, China}
\author{Dao-Xin Yao}
\email[]{yaodaox@mail.sysu.edu.cn}
\affiliation{State Key Laborator of Optoelectronic Materials and Technologies, School of Physics, Sun Yat-Sen University, Guangzhou 510275, China}
\author{Man-Rong Li}
\email[]{limanrong@mail.sysu.edu.cn}
\affiliation{Key Laboratory of Bioinorganic and Synthetic Chemistry of Ministry of Education, School of Chemistry, Sun Yat-Sen University, Guangzhou 510275, China}


\date{\today}

\begin{abstract}
Recent experimental work shows that FePS$_3$ undergoes phase transitions from $C2/m$ ($\beta\sim107^{\circ}$) to $C2/m$ ($\beta\sim90^{\circ}$) at $6$ GPa and then to metallic $P\bar{3}1m$ at $14$ GPa, with the magnetic ordering wave vector turning from $k=(01\frac{1}{2})$ to $k=(010)$ at $2$ GPa and to short-range magnetic order accompanying the insulator-metal transition. By preserving the magnetic point groups in $ab \ initio$ calculations we report the following: (1) We successfully reproduce the first magnetic structure transition at $1.2$ GPa and briefly discuss the influence of the Hubbard U parameter on this transition. This isostructural transition causes a change of the Brillouin zone from base-centered monoclinic to primitive monoclinic, and an indrect band gap to direct band gap transition. (2) There is a rotation of the Fe-S octahedron about $0.5^\circ$ through the $[001]$ axis before the neighboring layers shift. (3) The shift between neighboring layers is predicted to occur at $10.0$ GPa and reverses the energy order between $d_{x^2-y^2}$ and $d_{xy}$. (4) A sudden decrease of Fe-S bond length to $2.20$ \AA \  accompanies the vanishing of magnetic moment in the insulator-metal transition. Our work shows the importance of symmetries of magnetic structures in pressure-induced phase transition of magnetic systems.
\end{abstract}


\maketitle


\section{Introduction}

The $TM$P$X_3$ ($TM$ = Mn, Fe, Co, Ni, and Cd, $X$ = S, Se) layered compounds are ideal candidates for studying two-dimensional magnetism due to their diverse magnetic structures\cite{LeFlem1981} and weak coupling between layers. For example, at ambient pressure, the Mn$^{2+}$ ions couple antiferromagnetically along $a$ and $b$ for MnPSe$_3$, with magnetic moments lying in the $ab$ plane\cite{MnPSe3_1981}. The Mn$^{2+}$ ions also couple antiferromagnetically along $a$ and $b$ for MnPS$_3$, while the magnetic moment is perpendicular to the $ab$ plane\cite{MnPS3_2016}. For FePS$_3$, the Fe$^{2+}$ ions couple ferromagnetically along $a$ and are antiferromagnetically coupled along $b$. The magnetic moment is perpendicular to the $ab$ plane\cite{FePS3_2016} and can be maintained down to the few-layer\cite{Wang2016} or monolayer limit\cite{Lee2016}, as revealed by Raman spectroscopy.

The diversity of magnetic structures in this family of transition metal chalcogenophosphates is a combinational result of difference in magnetic anisotropy\cite{MPS3_1992}, number of electrons and the splitting of transition metal $d$ orbitals, and the overlapping between the transition metal $d$ orbitals and chalcogenide $p$ orbitals\cite{Coval_mag_2016}. The latter two determines the direct TM$^{2+}$-TM$^{2+}$ exchange and indirect TM$^{2+}$-$X^{2-}$-TM$^{2+}$ superexchange interaction, which are sensitive to the distance between TM$^{2+}$ and $X^{2-}$. Thus pressure\cite{Kim2019} is expected to tune the magnetic structure effectively, as recently demonstrated in MnP$X_3$ ($X$=S, Se)\cite{MnPX3_2016} and FePS$_3$\cite{FePS3_2019}, and offers a powerful method to find intriguing phases of matter\cite{PhysRevX.8.031059,PhysRevLett.117.146402}.

However, applying hydrostatic pressure often causes structural phase transition\cite{Jarosite_2020,Mn3TeO6_2019}, change of Brillouin zone and the site symmetry of the transition metal\cite{Liu2020}, making the analysis of magnetic structure difficult. Recent experimental works\cite{Haines2018,FePS3_2021} on FePS$_3$ show that at low pressure (less than $6$ GPa) it undergoes an isostructural transition, preserving the $C2/m$ space group, while the stacking behavior is changed, with $\beta$ turning from $107^{\circ}$ to about $90^{\circ}$. Accompaning this transition is a change of magnetic structure, with magnetic propagation vector turning from $\vec{k}=(01\frac{1}{2})$ to $\vec{k}=(010)$. At higher pressure (about $14$ GPa) it further transforms to trigonal structure with space group $P\bar{3}1m$ and becomes metallic.

 Based on the Haines \textit{et al.}'s experimental work\cite{Haines2018}, by using LDA+U (with U equal to $2.5$ eV) method and comparing four kinds of structures ($C2/m(\beta\sim107^\circ)$, $C2/m(\beta\sim90^\circ)$, $P\overline{3}1m$ and $R\overline{3}$) and their relative stabilities, Zheng \textit{et al.}\cite{FePS3_cal_2019} reproduced the isostructural phase transition from $C2/m(\beta\sim107^\circ)$ to $C2/m(\beta\sim90^\circ)$ at about $5$ GPa, with zigzag antiferromagnetic ground state, namely $k=(010)$. The $C2/m(\beta\sim90^\circ)$ phase is predicted to turn into $P\overline{3}1m$ at $17$ GPa. The magnetic moment of Fe was found to vanish in the metallic $P\overline{3}1m$ phase. The magnetic structure transition from $k=(01\frac{1}{2})$ to $k=(010)$ were not studied, and the vanishing of magnetic moment was in conflict with the short-range magnetic structure, both are revealed by more recent experimental work by Coak \textit{et al.}\cite{FePS3_2021}. By first-principle linear combination of atomic orbitals calculations, Evarestov \textit{et al.}\cite{FePS3_cal_2020} studied the origin of pressure-induced insulator-metal transition in FePS$_3$. However, they considered only nonmagnetic structures and hybrid DFT-HF functionals were used, thus the magnetic phase transition were not discussed there.

To consider the detailed magnetic structure in \textit{ab initio} calculations is challenging. Sometimes it's due to the lack of magnetic structure data from experiment. In other cases, the magnetic structure is incommensurate with the paramagnetic crystal structure and a large supercell has to be used in order to incorporate the magnetic structure. Luckily, the magnetic interaction energy is often in $meV$ order of magnitude and can be considered as perturbation compared with the enthalpy change in structural phase transition. This fact suggests that it might give reliable result about phase transition if we use a commensurate magnetic structure that is closest to the original one in the calculation.

Following this thought, here we study the crystal, magnetic and electronic structures of FePS$_3$ under pressure up to $50$ GPa, by preserving the experimental magnetic point group structure in $ab \  initio$ calculations. Not only are the two phase transitions found by Zheng \textit{et al.}\cite{FePS3_cal_2019} reproduced here, we also identify that (1) the low pressure isostructural phase transition takes three steps to finish: at about $1.2$ GPa the magnetic propagation vector changes from $k=(01\frac{1}{2})$ to $k=(010)$, then there is a rotation of the Fe-S octahedron about $0.5^\circ$ through $[001]$ at $6.0$ GPa, and the shift of neighboring layers along $a$ happens at $10.0$ GPa. (2) The splitting of $d$ orbitals of Fe is closely related to the Fe-S bond length and the shift of neighboring layers can reverse the energy order of $d_{xy}$ and $d_{x^2-y^2}$. (3) The change of Brillouin zone and the band structure due to the magnetic and crystal structure transition are reported. (4) The discrepancy between the predicted vanishing of magnetic moment and the experimental finding of short-range magnetic order in the metallic phase is discussed. In section \ref{sec:method} the methods of calculations are given, and the results and discussions are presented in section \ref{sec:results}. A brief conclusion is given in section \ref{sec:conclusion}.

\section{Methods}
\label{sec:method}
The $ab \  initio$ calculations are performed using the Vienna \textit{ab initio} package (VASP)\cite{VASP}, with projector augmented-wave (PAW)\cite{Blochl_PAW} method to construct the pseudopotentials. The cutoff energy is set to $1000$ eV, and a Monkhorst-Pack grid\cite{MP_1976} of $11\times11\times9$ ($7\times7\times11$ for magnetic propagation vector $k=(01\frac{1}{2})$) for integration of $k$ points is used. The criterion of convergence of total energy and forces are $10^{-9}$ eV and $0.01$ eV/\AA, respectively. To depict the van der Waals interaction between layers the DFT-D2 method of Grimme\cite{Grimme_2006} is used. The LDA+U method introduced by Dudarev $et al.$\cite{Dudarev_1998} is adopted, due to the localized $d$ electrons of Fe, with $U=4.0$ eV. The energy-volume (E-V) curves are calculated by fixing the magnetic point group symmetry and the volume of the primitive cell during optimization, while the atoms are fully relaxed until convergence on forces is reached.

The phonon spectrum is calculated based on the supercell method by using phonopy\cite{phonopy}. A $2\times2\times2$ supercell is used and an cutoff energy of $700$ eV and $5\times5\times5$ Monkhorst-Pack grid are adopted in the force calculations.

\begin{figure}[t]
	\includegraphics[width=8cm]{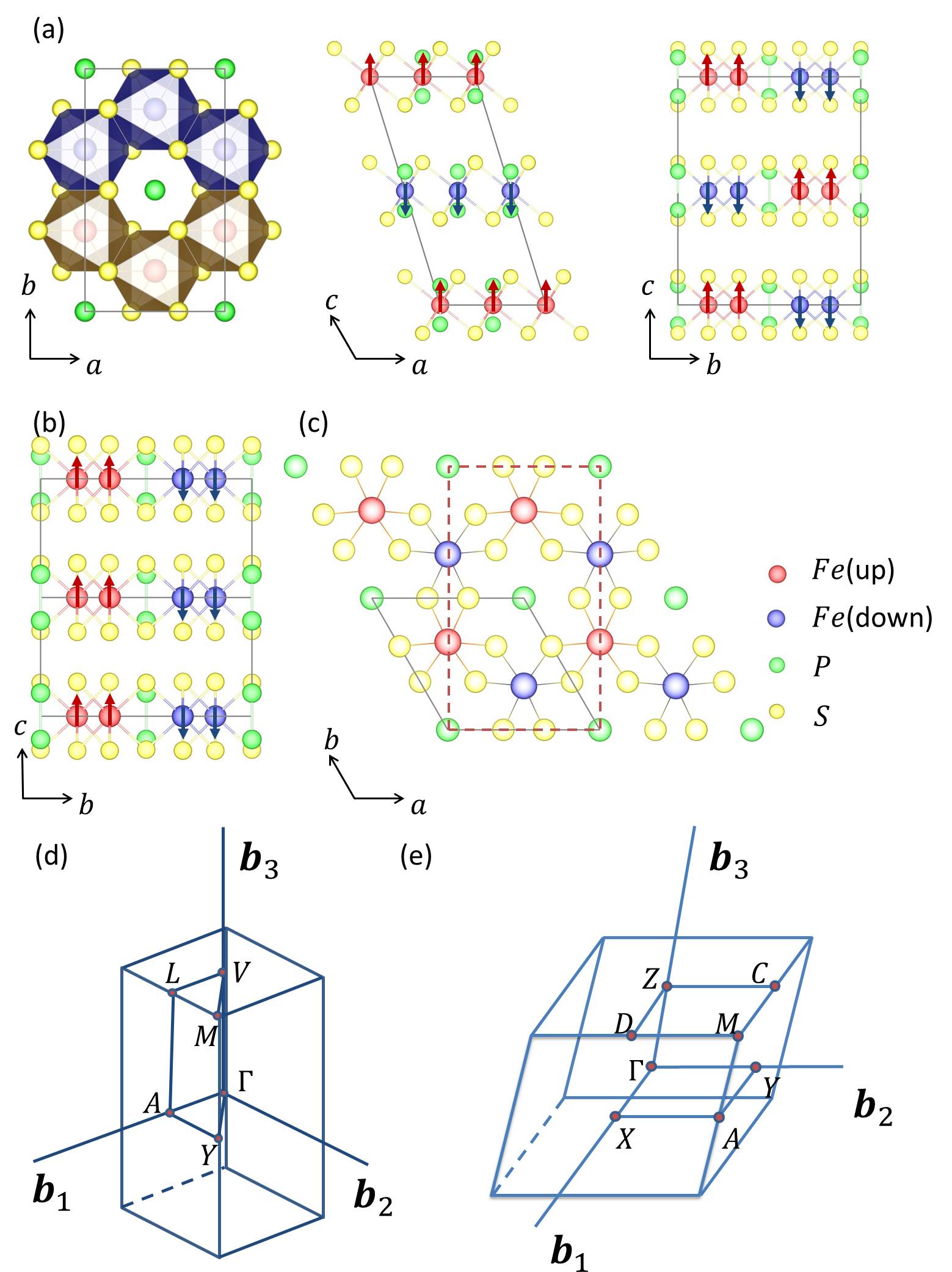}
	\caption{The crystal structure of FePS$_3$ in (a) $C2/m(\beta\sim107^\circ)$ with magnetic propagation vector $k=(01\frac{1}{2})$, (b) $C2/m(\beta\sim90^\circ)$ with $k=(010)$ and (c) $P\bar{3}1m$ with $k=(000)$. The left panel in (a) shows the magnetic structure in $ab$ plane, and the middle and right panels correspond to the magnetic structure in $ac$ and $bc$ planes, respectively. In (c) the unit cell of $C2/m$ is also shown in dashed rectangle. (d) and (e) show the Brillouin zones of $C2/m(\beta\sim107^\circ,k=(01\frac{1}{2}))$ and $C2/m(\beta\sim107^\circ,k=(010))$, which are base-centered monoclinic and primitive monoclinic, respectively\cite{BZ}. }
	\label{fig:structure}
\end{figure}

\section{Results and Discussion}
\label{sec:results}

According to experimental results, FePS$_3$ crystallizes in monoclinic $C2/m$\cite{MPX3_1985} at ambient pressure and  trigonal $P\bar{3}1m$\cite{Haines2018} space groups at high pressure, which have similar intralayer structure but different interlayer stacking. In each layer the Fe$^{2+}$ ion is surrounded by S octahedron and forms a honeycomb network by sharing octahedron edges with nearest neighbors. The P-P 'molecule' locates in the honeycomb center. The Fe$^{2+}$ ion has three nearest neighbors Fe$^{2+}$ within the layer for both $C2/m$ and $P\bar{3}1m$, two nearest neighbors with each adjacent layer for $C2/m (\beta\sim107^{\circ})$, while one nearest neighbors with each adjacent layer for $C2/m (\beta\sim90^{\circ})$ and $P\bar{3}1m$, respectively (Fig.\ref{fig:structure}(a-c)).

\begin{figure}[t]
	\includegraphics[width=8cm]{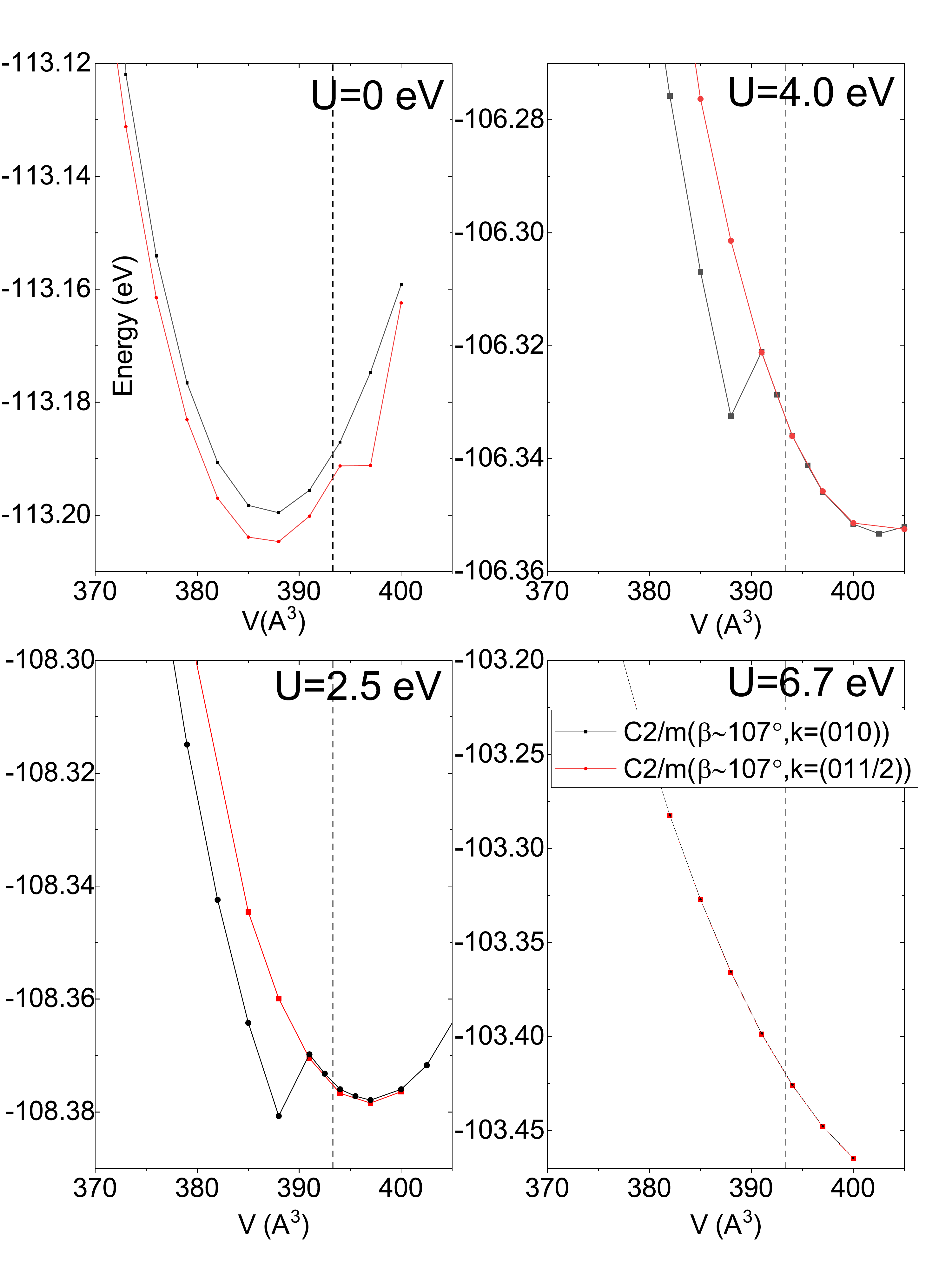}
	\centering
	\caption{The effect of $U$ on the E-V curves of $C2/m(\beta \sim 107^\circ,(010))$ and  $C2/m(\beta \sim 107^\circ,(01\frac{1}{2}))$ near V=$388$ \AA$^3$. The vertical dashed line indicates the experimental volume at ambient pressure.}
	\label{fig:EVU}
\end{figure}

\begin{figure}[t]
	\includegraphics[width=8cm]{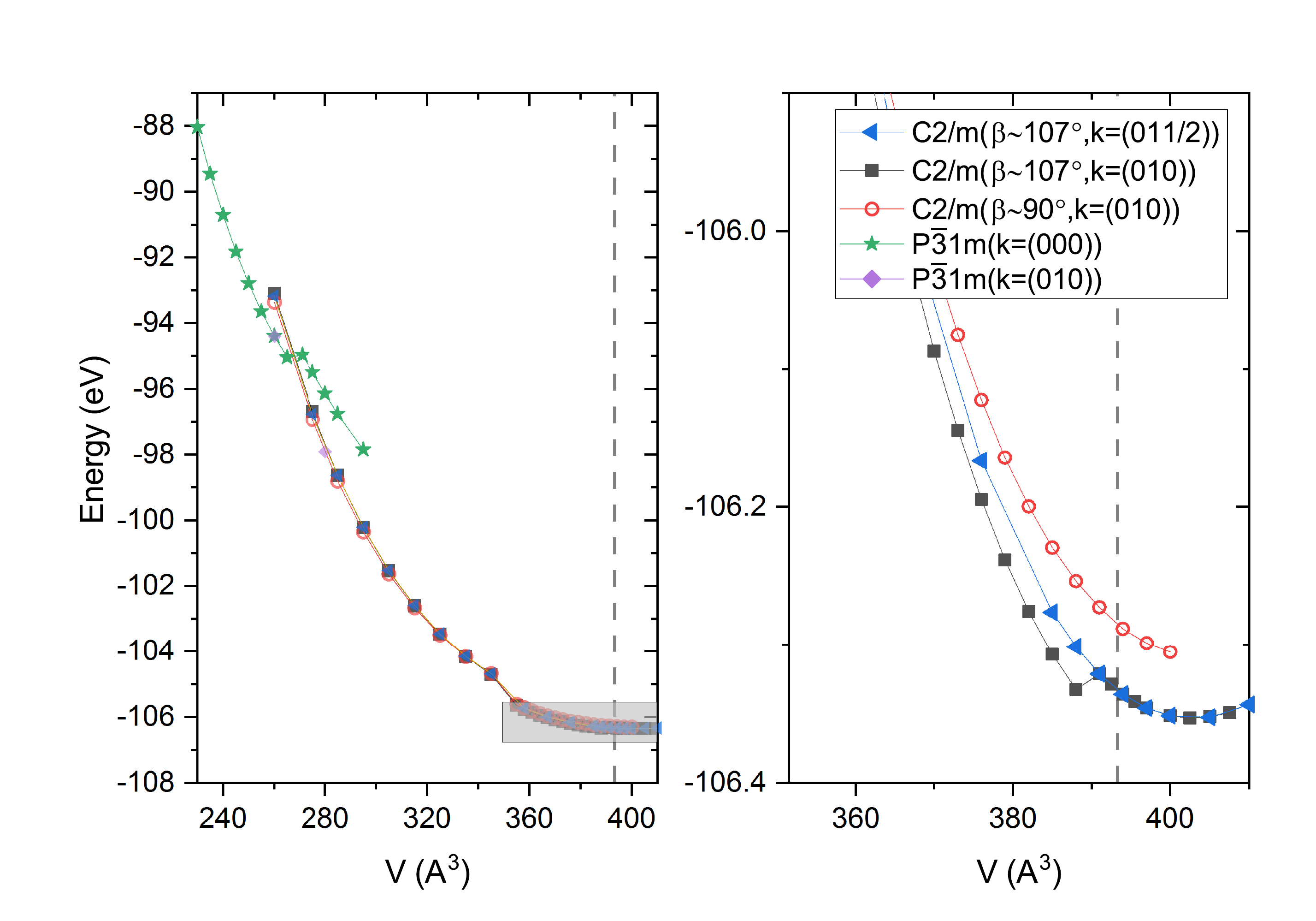}
	\caption{The $E-V$ curves of FePS$_3$. The vertical dashed line represents the experimental value of the unit cell volume at ambient pressure. The right panel is the enlarged view of the rectangular region of the left panel. Due to discontinuity at $V=350$ \AA$^3$, the E-V curves for $C2/m(\beta \sim 107^\circ, k=(01\frac{1}{2}))$, $C2/m(\beta \sim 107^\circ, k=(010))$ and $C2/m(\beta \sim 90^\circ, k=(010))$ are divided into 'L' (low pressure) and 'H' (high pressure) for $V > 350$ and $V<350$ \AA$^3$, respectively. Note that for $C2/m(\beta \sim 107^\circ, k=(010))$, there is another discontinuity point at $V=390$ \AA$^3$, and $C2/m(\beta \sim 107^\circ, k=(010))L$ is thus confined in the region $350<V<390$ \AA$^3$. The discontinuity at $V=270$ \AA$^3$ divides the E-V curve for $P\overline{3}1m$ into `L' and `H' for $V>270$ and $V<270$ \AA$^3$, respectively.}
	\label{fig:EV}
\end{figure} 

\begin{figure}[t]
	\begin{minipage}[b]{\linewidth}
		\centering
		\subfigure[]{\includegraphics[width=8cm]{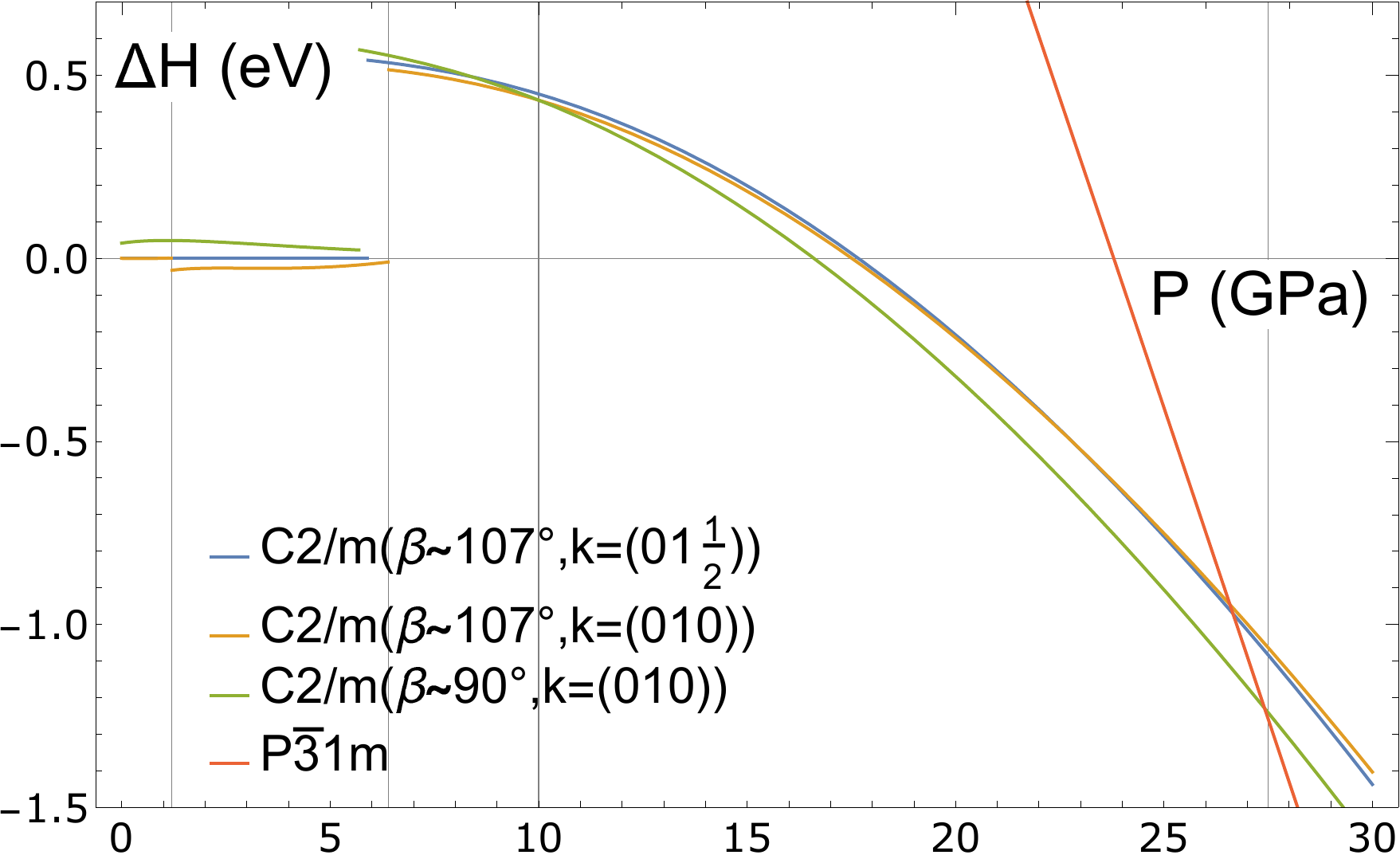}}
	\end{minipage}
	\medskip
	\begin{minipage}[b]{\linewidth}
		\centering
		\subfigure[]{\includegraphics[width=4cm]{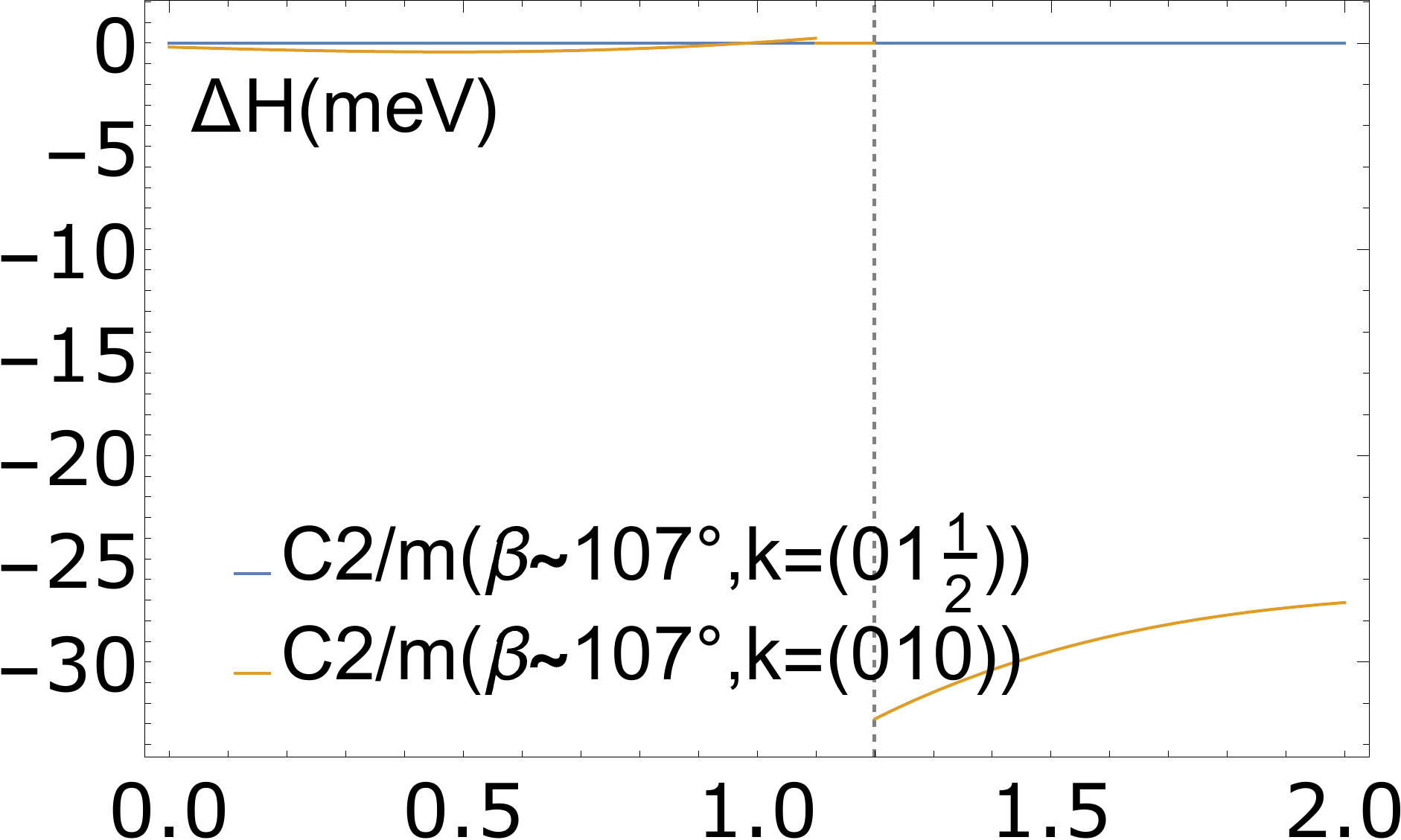}}
		\subfigure[]{\includegraphics[width=4cm]{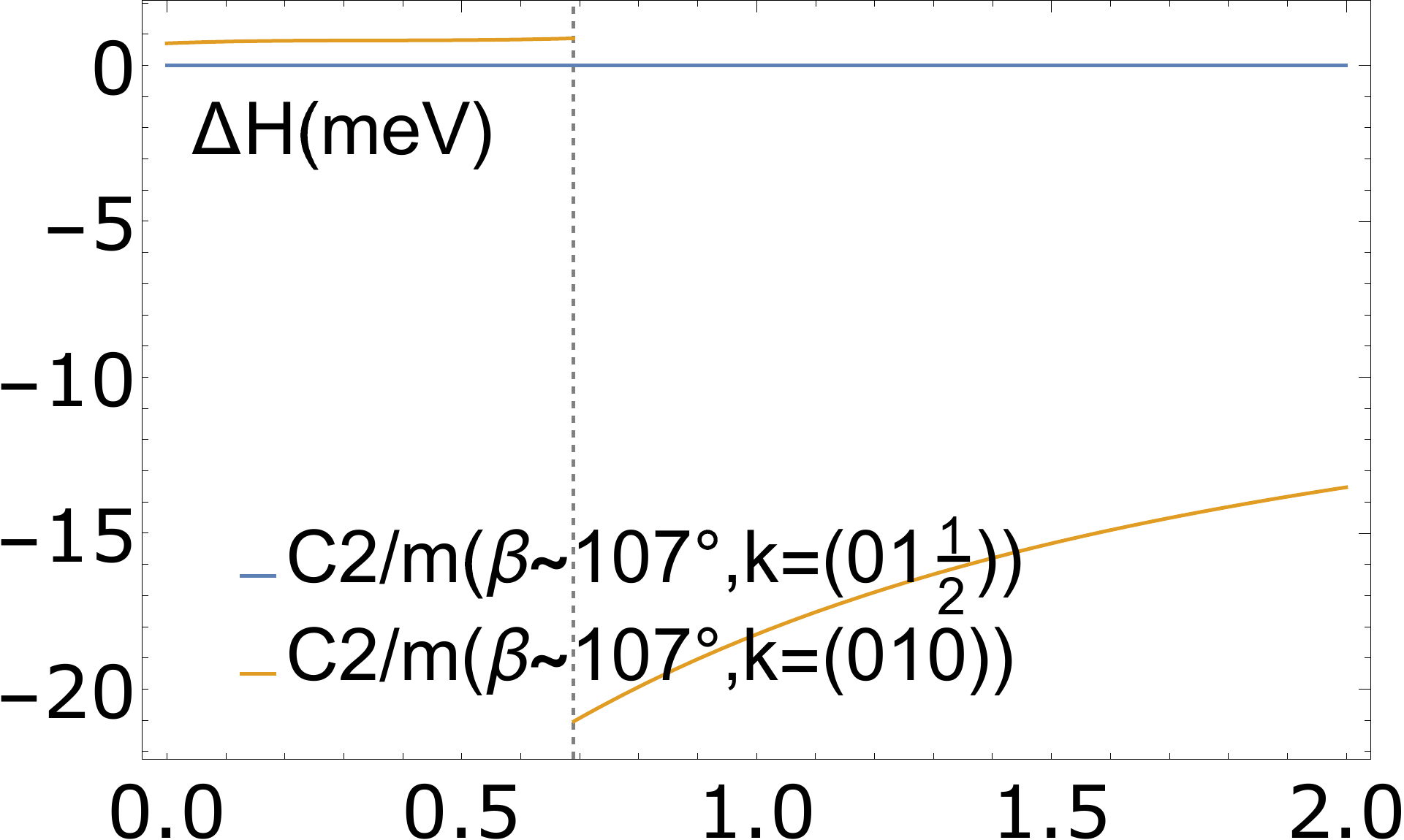}}
	\end{minipage}
	\caption{(a) The enthalpy $H$ as a function of pressure $P$ of FePS$_3$. The enthalpy of the $C2/m(\beta \sim 107^\circ,k=(01\frac{1}{2}))L$ structure is taken as the reference. The vertical dashed lines indicate the transition pressures. Note that the discontinuities of enthalpy of $C2/m(\beta \sim 107^\circ,k=(010))$ near $1.0$ GPa and all the $C2/m$ phases near $6.0$ GPa result from the discontinuity of E-V curves at the corresponding volumes. (b) and (c) are the $H-P$ curves near $1.0$ GPa with $U=4.0$ and $U=2.5$ eV, respectively.}
	\label{fig:HP}
\end{figure}

\subsubsection{Influence of U on predicted phase transition}
LDA+U improves the description of systems with localized $d$ electrons, by introducing a strong intra-atomic interaction. However, $U$ depends on the local chemical environment of the Hubbard site\cite{DFTU1,DFTU2}, and it is crucial to choose a suitable U to get reasonable results about phase transitions in antiferromagnetic systems like FePS$_3$. Here we take the $C2/m (\beta\sim 107^\circ,k=(01\frac{1}{2}))$ to $C2/m(\beta\sim 107^\circ,k=(010))$ phase transition to illustrate the influence of U.  By comparing the $E-V$ curves (see Fig.\ref{fig:EVU}) of the two phases with different $U$, we find that (a) The equilibrium volume $V_0$ increases as U is increased, because a larger $U$ (on-site Coulomb repulsion)) tends to decrease the charge density in interstitial region, and reduce the covalent bonding strength \cite{Dudarev_1998}. (b) $U=0$ and $U=6.7$ eV cannot predict the $k=(01\frac{1}{2})$ to $k=(010)$ phase transition, although they predict the correct ground state. Besides, $U=0.0$ gives metallic ground state at ambient pressure (see Fig.S3  and Fig.S4). (c) Both $U=2.5$ and $U=4.0$ can predict the anomaly of E-V curves of $k=(010)$ near $V=388$ \AA$^3$, while $U=4.0$ gives nearly the same $E-V$ curves for $k=(01\frac{1}{2})$ and $k=(010)$ for $V>388$ \AA$^3$, with the energy difference at the same volume less than $1$ meV. This also affects the H-P curve given below. We choose $V=394$ \AA$^3$ and calculate the energy difference between  $C2/m (\beta\sim 107^\circ,k=(01\frac{1}{2}))$ and $C2/m(\beta\sim 107^\circ,k=(010))$ (see Fig.S2) and find that energy difference varies with U in a nonlinear way\cite{SM}. A full discussion of this effect is beyond the scope of the present work.

The calculated $E-V$ curves indicate there are four phase transitions in the volume between $400$ and $240$ \AA$^3$ (Fig.\ref{fig:EV}). The transition pressure, difference between phases and the stable pressure region of each phase will be given soon. Each $E-V$ curve is obtained by constraining the primitive unit cell to a certain space group and magnetic structure, as indicated in the legend. Note that the total energy of each phase, as a function of volume, is piecewise continuous. Discontinuity of a $E-V$ curve (e.g. $C2/m(\beta \sim 107^\circ, k=(010))$ at $V=350$ and $388$ \AA$^3$, $P\overline{3}1m$ at $V=270$ \AA$^3$) implies a hidden discontinuity of physical quantity, often the crystal structure, magnetic moment, or conductivity, which we will clarify later. For each continuous region the Murnaghan equation\cite{Murnaghan1944} is applied to fit the $E-V$ data to get equation of state, and the relation between volume and pressure $V(P)$ can be determined by  $P=-dE/dV$. The fitted parameters are summarized in Table \ref{tab:Bulk_modulus}. In this way, the enthalpy of each phase can be written as a function of pressure, $H=E(V)+PV=E(V(P))+P\times V(P)$, the pressure dependence of the enthalpy (Fig.\ref{fig:HP}), lattice constant and magnetic moment are derived. The phase with lowest enthalpy is the ground state phase, and the transition pressure can be determined from the $V-P$ curve of the low pressure phase at the transition volume, whether it is located at the discontinuity point of E-V curve or not.

In this way, we reach the following conclusion about transition pressures: (1) The ambient pressure structure  $C2/m(\beta\sim107^\circ)L$ undergoes an isostructural phase transition with magnetic propagation vector transformed from $k=(01\frac{1}{2})$ to $k=(010)$ at $0.7$ GPa with $U=2.5$ eV (Actually, due to the energy difference between $k=(01\frac{1}{2})$ and $k=(010)$ is smaller than $1$ meV, we cannot tell that $k=(01\frac{1}{2})$ is the ground state when $P<1.2$ GPa with $U=4.0$ eV, see Fig.\ref{fig:HP} (b) and (c).); (2) The ambient pressure structure $C2/m(\beta\sim107^\circ)L$ turns into $C2/m(\beta\sim107^\circ)H$ at about $6.0$ GPa (The meaning of `$L$' and `$H$' is given in the caption of Fig.\ref{fig:EV}); (3) The $C2/m(\beta\sim107^\circ)H$ structure turns into the $C2/m(\beta\sim90^\circ)H$ structure at about $10.0$ GPa; The magnetic propagation vector is $k=(010)$ in (2) and (3); (4) The $C2/m(\beta\sim90^\circ)H$ structure turns into the nonmagnetic $P\bar{3}1m$ structure at about $27.5$ GPa. We now expound these phase transitions in detail.

\subsubsection{Structural analysis}

\textit{Phase transition (1)}. There is a large lattice distortion accompanying the phase transition (1). The most obvious change is that Fe-Fe bond along $b$ is reduced from $3.57$ to $3.35 $ \AA, which may be reflected by Raman spectroscopy as the 'stiffness' between the Fe atoms is changed abruptly. Note that this bond is between iron atoms of opposite magnetic moments (spins), and the bond length of iron atoms of same magnetic moments (spins) gets larger, from $3.37$ to $3.47$ \AA. Also note that this sudden change of bond length exists only in $C2/m(\beta\sim107^\circ)$ with $k=(010)$ and is associated with an energy jump about  $30$ meV. For $k=(01\frac{1}{2})$ there is no discontinuity around $V=388$ \AA$^3$ and no such change of Fe-Fe bond length, indicating that interlayer magnetic interactions is non-neglible. Similar sudden changes of bond length under pressure-induced phase transition is observed in MnTa$_2$O$_6$\cite{Liu2020}. Another interesting feature is that the calculated bulk modulus at ambient pressure is reduced after the phase transition, turning from $26.2$ GPa at ambient pressure to about $9.4$ GPa (see Table.\ref{tab:Bulk_modulus}). This `anomaly' occurs because the ambient pressure volume of $C2/m(\beta\sim107^\circ,(010))L$ is larger than that of $C2/m(\beta\sim107^\circ,(01\frac{1}{2}))L$.

\begin{figure}[t]
	\includegraphics[width=8cm]{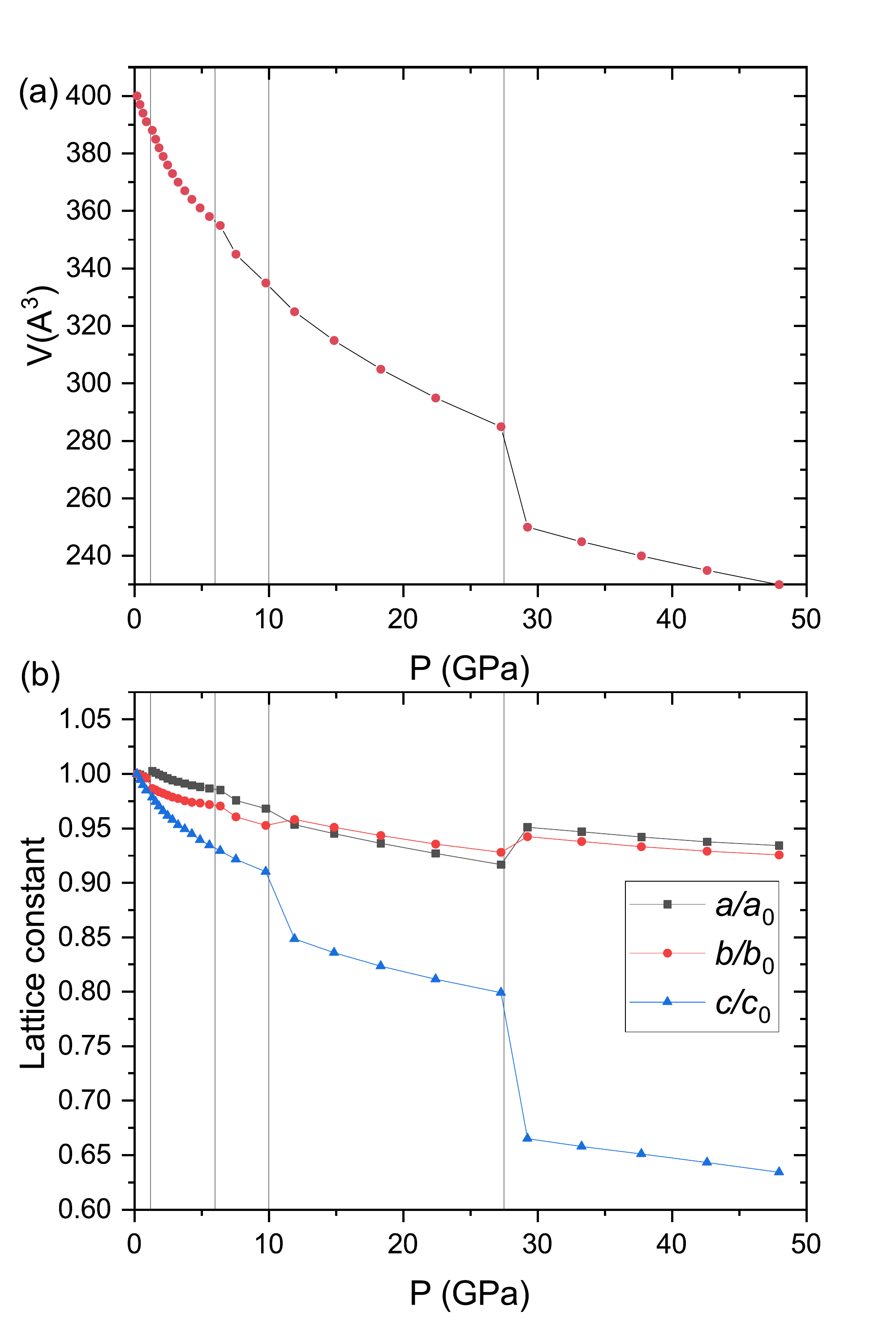}
	\caption{ (a) The unit cell volume of FePS$_3$ as a function of pressure. For $P\geq27.5$ GPa the unit cell volume is that of a supercell of the hexagonal lattice formed by $\vec{t_1}$, $2\vec{t_2}-\vec{t_1}$, and $\vec{t_3}$.  (b) The pressure dependence of lattice constants of FePS$_3$. Here the $a_0$, $b_0$ and $c_0$ are the lattice constants of $C2/m(\beta\sim107^\circ,(01\frac{1}{2}))$ at volume of $400$\AA. For $P\geq27.5$ GPa $b=\sqrt{3}a$. The vertical lines indicate the predicted phase transition pressures.}
	\label{fig:VP}
\end{figure}

\textit{Phase transition (2)}. An isotructural phase transition that is not observed in experiment. The space group remains as $C2/m(\beta\sim107^\circ)$ and the lattice constants do not exhibit discernable discontinuities. What's unusual here is that the other two metastable structures ($C2/m(\beta\sim90^\circ,k=(010))$ and $C2/m(\beta\sim107^\circ,k=(01\frac{1}{2}))$) also show discontinuity at the transition pressure, as shown in  Fig.\ref{fig:EV}, where three curves are discontinuous near $V=355$ \AA$^3$. Note that this phase transition should not be considered as artificial due to neglecting of zero-point energy, as the phase transition is isostructural with lattice constant being almost unchanged. The phonon spectrum and thus zero-point energy should be almost the same, and neglecting it would not result in the phase transition. Rather, this predicted transition is found to be associated with a rotation about $0.5^\circ$ of the FeS$_6$ octahedron about the $[001]$ axis. As shown in the set of Fig.\ref{fig:mu_P}(b), in the six octahedron forming the hexagon in the $ab$ plane, half rotate clockwise and the other half anticlockwise. This deformation increases the total energy of the crystal quickly and, after a certain angle, the increase in total energy would be higher than that caused by shifting between layers, which is phase transition (3). 

\textit{Phase transition (3)}. This phase transition is also isostructural in that the space group $C2/m$ is not changed. However, the beta angle changes drastically from about $107^\circ$ to about $90^\circ$ (See Fig.\ref{fig:structure}(a) and (b)). This transition occurs by shifting the upper FePS$_3$ layer $\frac{1}{3}a$ along the $a$ direction, relative to the lower layer. Accompanying this shift is a drop of the $c$ lattice constant (Fig.\ref{fig:VP}(b)). This transition is observed by experiment, though, at a lower pressure ($~\sim6.0$ GPa).

\textit{Phase transition (4)}. The $C2/m(\beta \sim 90^\circ)H$ structure transforms into the $P\bar{3}1m$ structure at $27.3$ GPa, which is larger than the experimental value $14$ GPa. There are two reasons for this discrepancy, one is the neglect of zero-point energy of the two involved structures\cite{Kim2019}, another reason will be clarified later. The volume collapse is about $10\%$, with the $c$ lattice constant reduced significantly. Examination of the crystal structure shows that after the transition, the FeS$_6$ octahedron is compressed to such an extent that the Fe-S bond length is reduced to $2.18$ \AA \  (This bond length is $2.54$ \AA \   in ambient pressure), even smaller than the sum of Fe$^{2+}_{LS}$ and $S^{2-}$ ionic radii (see Fig.\ref{fig:mu_P}(a)). This causes large overlap between the Fe $3d$ and the S $3p$ orbitals. This overlap, together with the $D_{3d}$ point group symmetry, causes the metallic behavior discussed below. Similar Fe-S bond length about $2.20$ \AA \  were also found in BaFe$_2$S$_3$ under pressure\cite{PhysRevB.98.180402}.

\begin{table}
	\caption{The bulk modulus $B_0$ at ambient pressure, the first derivative of bulk modulus $B_0'$ with respect to pressure $B_0'=\partial B/\partial P$ at ambient pressure and equilibrium volume $V_0$ of FePS$_3$ in each phase, in each continuous region.}
	\begin{tabular}{lp{1.5cm}p{1.5cm}l}
		\hline
		Structure & $B_0$ (GPa) & $B_0'$ &  $V_0$ (\AA$^3$) \\
		\hline
		$C2/m(\beta\sim107^{\circ},(01\frac{1}{2}))L$ & $26.2\pm0.5$ & $8.4\pm0.3$ & $403.0\pm0.1$ \\
		$C2/m(\beta\sim107^{\circ},(01\frac{1}{2}))H$ & $45.5\pm3.6$ & $3.6\pm0.2$ & $392.7\pm3.5$ \\
		$C2/m(107^{\circ},(010))V>388\AA^3$ & $26.2\pm1.0$ & $10.5\pm3.4$ &  $402.9\pm0.1$ \\
	    $C2/m(\beta\sim107^{\circ},(010))L$ & $9.4\pm2.8$ & $14.2\pm0.9$ &  $419.5\pm5.4$ \\
	    $C2/m(\beta\sim107^{\circ},(010))H$ & $42.0\pm0.9$ & $4.0\pm0.1$ &  $395.2\pm0.8$ \\
	    $C2/m(\beta\sim90^{\circ},(010))L$ & $20.8\pm1.1$ & $9.9\pm0.5$ &  $405.7\pm0.7$ \\
	    $C2/m(\beta\sim90^{\circ},(010))H$ & $41.1\pm0.8$ & $4.0\pm0.1$ &  $393.9\pm0.7$ \\
	    $P\bar{3}1m(k=(000))H$ & $89.8\pm5.2$ & $3.5\pm0.2$ &  $310.5\pm1.7$ \\
		\hline		
	\end{tabular}
	\label{tab:Bulk_modulus}
\end{table}

\subsubsection{Electronic structure}

We now proceed to analyze the variation of electronic structures of FePS$_3$ with pressure. At ambient pressure, the magnetic propagation vector is $k=(01\frac{1}{2})$, this causes a doubling of the unit cell along the $c$ direction. However, the primitive cell is a centered monoclinic in $bc$ plane (Fig.\ref{fig:structure}(a)), and the volume of the Brillouin zone is the same as that of a primitive monoclinic, only the shape is different. After phase transition (1), the magnetic propagation vector is $k=(010)$, and the primitive cell is a primitive monoclinic. Thus, phase transition (1) changes the Brillouin zone from centered monoclinic to primitive monoclinic, as shown in Fig.\ref{fig:structure}(d) and (e).

From the magnetic space group point of view, the ambient pressure magnetic structure has the magnetic space group $C_c2/m$, which preserves the inversion symmetry $\{-1|0\}$ and has the additional $T\{0|\frac{1}{2}\frac{1}{2}0\}$ symmetry. In the $k=(010)$ phase, the magnetic space group is $P_c2_1/m$, in which both the inversion symmetry and time reversal symmetry are broken, while the combined symmetry $T\{-1|0\}$ is preserved.

\begin{widetext}
	\begin{figure*}[t]
		\includegraphics[width=16cm]{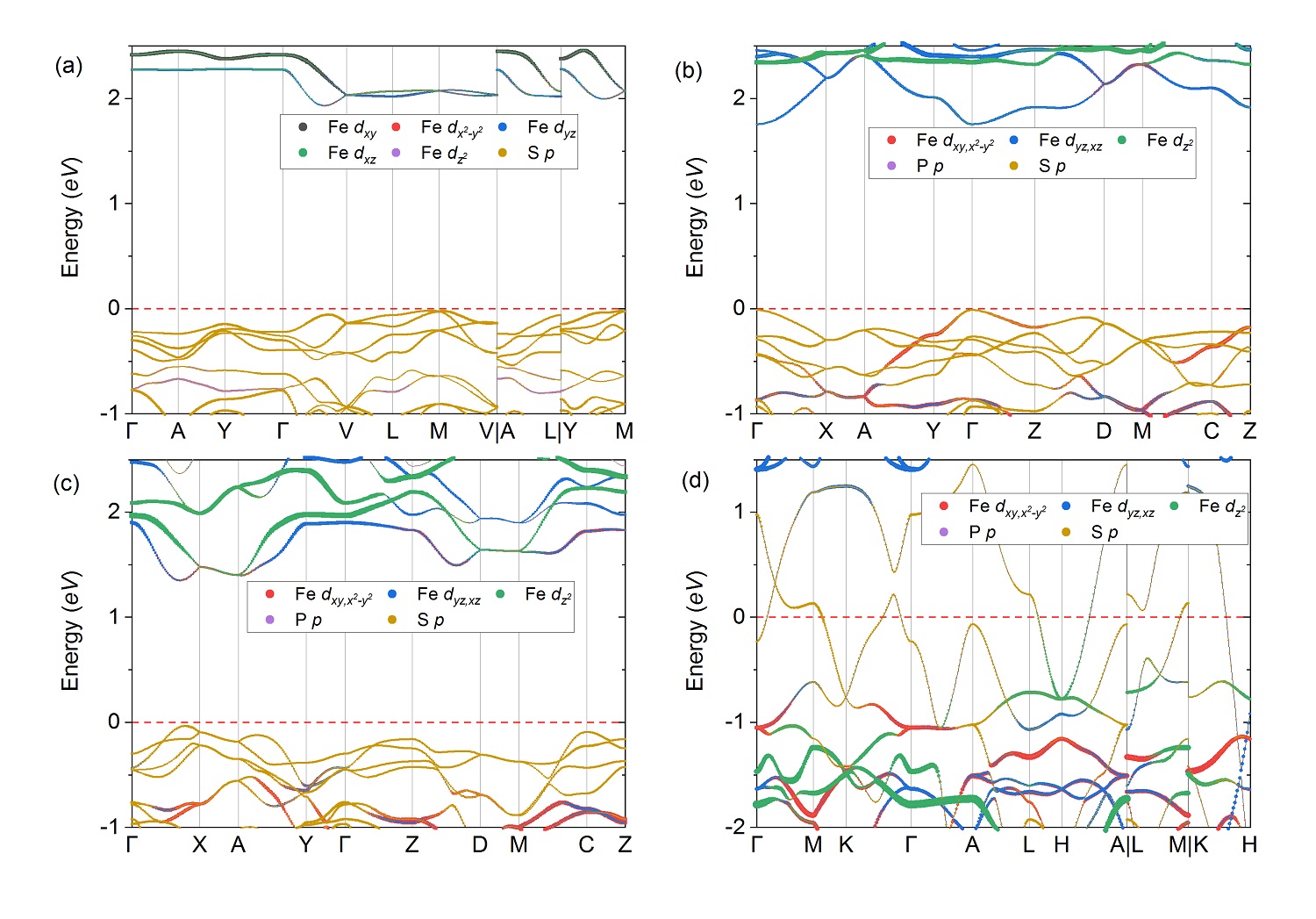}
		\caption{The band structures of FePS$_3$ in (a) $C2/m (\beta\sim 107^\circ,k=(01\frac{1}{2}))$ (ambient pressure), (b) $C2/m(\beta\sim 107^\circ,k=(010))$ ($P=2.1$ GPa), (c) $C2/m(\beta\sim 90^\circ,k=(010))$ ($P=14.9$ GPa) and (d) $P\overline{3}1m$ ($P=29.3$ GPa) structures, respectively. }
		\label{fig:BS}
	\end{figure*}
\end{widetext}

At ambient pressure, FePS$_3$ shows an indirect band gap about $1.95$ eV\cite{FePS3_bs_2016}, with valence band maximum (VBM) located along $ML$ and originated from S $3p$ + Fe $3d_{x^2-y^2}$ orbitals, and conduction band minimum (CBM) located along $\Gamma V$ and originated from Fe $3d$ orbitals (see Fig.\ref{fig:BS}(a) and Fig.S5). After phase transition (1), it turns into a direct band gap semiconductor, with the band gap reduced to $1.77$ eV and both CBM and VBM located at $\Gamma$. The CBM and VBM originate again from Fe $3d$ and S $3p$ + Fe $d_{xy,x^2-y^2}$ states, respectively (Fig.\ref{fig:BS}(b)). The phase transition (2) has not much influence on the band structure, except that the band gap is reduced to $1.51$ eV. After phase transition (3), the neighboring layers are almost A-A stacking, and the CBM and VBM move to $\Gamma X$, with CBM originated again from Fe $3d$ orbitals, while VBM now comes from S $3p$ + Fe $d_{xz,yz}$ orbitals (Fig.\ref{fig:BS}(c)). Another change accompanying phase transition (3) is that the $d_{x^2-y^2}$ becomes higher in energy than that of $d_{xy}$ (Compare the relative positions of the DOS of $d_{x^2-y^2}$ and $d_{xy}$ in the spin minority channel, as shown in Fig.S6 and Fig. S7.). This change arises as the bond length of Fe-Fe bond along $b$ becomes larger than that along $[1\overline{1}0]$, as is obvious from the pressure dependence of lattice constants of $a$ and $b$ (Fig.\ref{fig:VP}(b)), where $b/b_0$ becomes larger than $a/a_0$ after phase transition (3). The Fe-Fe bond along $b$ has large overlap between $d_{x^2-y^2}$ orbitals while Fe-Fe bond along $[1\overline{1}0]$ has large overlap between $d_{xy}$ orbitals. The elongation of Fe-Fe bond along $b$ weakens the overlap between $d_{x^2-y^2}$ orbitals, which resembles dangling bond in some extent and thus becomes higher in energy.

\begin{figure}[h]
	\includegraphics[width=8cm]{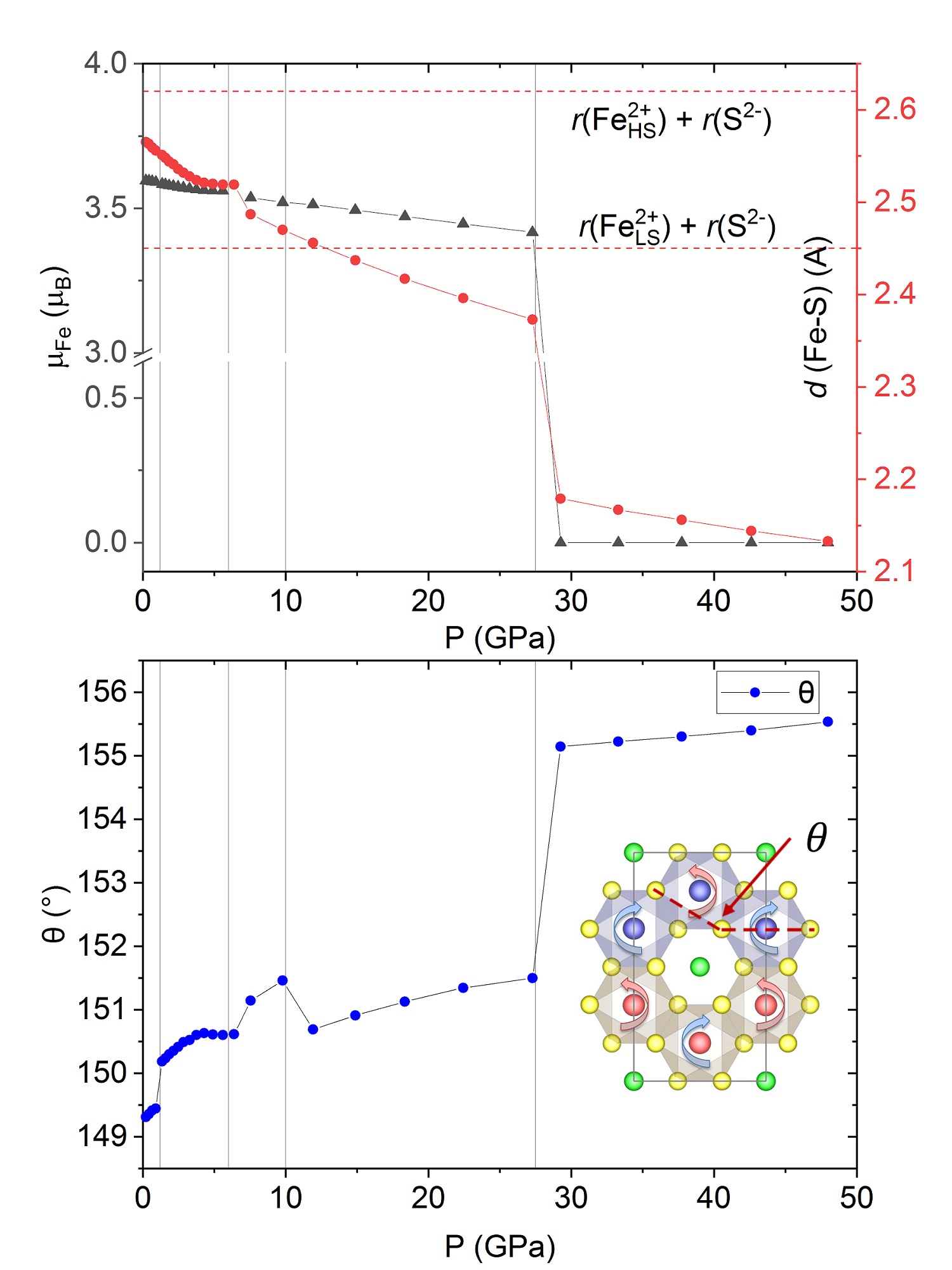}
	\caption{ (a) The magnetic moment of Fe$^{2+}$ ion and the Fe-S bond length as a function of pressure. The red dashed lines indicate the sum of ionic radius of Fe$^{2+}$ in octahedral coordination and S$^{2-}$,  in high-spin (HS) and low-spin (LS) states, respectively. (b)  The rotation of FeP$_6$ octahedron through $[001]$ as a function of pressure. The definition of $\theta$ and the rotations of the octahedron are  shown in the inset. }
	\label{fig:mu_P}
\end{figure}

After phase transition (4), FePS$_3$ shows metallic band struture, the Fermi surface is mainly composed of S $p$ + Fe $d_{z^2}$ orbitals (Fig.\ref{fig:BS}(d)). Furthermore, the magnetic moment of Fe vanishes (see Fig.\ref{fig:mu_P}(a)), which is consistent with the result of Zheng \textit{et al}. \cite{FePS3_cal_2019}. This vanishing of magnetic moment is reflected in the discontinuity of E-V curve for $P\overline{3}1m(k=(000))$ near $V=270$ \AA$^3$, as shown in Fig.\ref{fig:EV}. We observe that firstly, FePS$_3$ in $P\overline{3}1m$ structure at larger volume also shows metallic behavior but the magnetic moment of Fe does not vanish; second, FePS$_3$ in $C2/m$ with the same volume (smaller than $270$ \AA$^3$) as that of $P\overline{3}1m$ shows insulating behavior, with nonvanishing magnetic moment. The vanishing of magnetic moment of Fe gives rise to a decrease in energy, about $250$ meV per formula unit (see Fig.\ref{fig:EV} at $V=260$ \AA$^3$). Similar vanishing of magnetic moment of Fe has also been found in BaFe$_2$S$_3$\cite{PhysRevB.98.180402} and BaFe$_2$Se$_3$\cite{BaFe2Se3_2017} during pressure-induced insulator-metal transition, where Fe is surrounded by S tetrahedron, though, not octahedron.

It should be noted from  above analyses that the P $2p$ orbitals do not show significant contribution to the insulator-metal transition accompanying phase transition (4), as was suggested in Ref \cite{FePS3_2021}. Although in $P\overline{3}1m$ the P atoms has nearly the same $z$ coordinate as that of S atoms due to the compression of Fe-S octahedron, and P-P distance between layers are reduced to be the same as that in the layer, the P-P bond length (about $2.2$ \AA) does not show any decrease during the pressure range studied, perhaps due to the fact that neighboring high valence ions prefer to displace away from each other as a result of electrostatic repulsion.

\begin{figure}[h]
	\includegraphics[width=8cm]{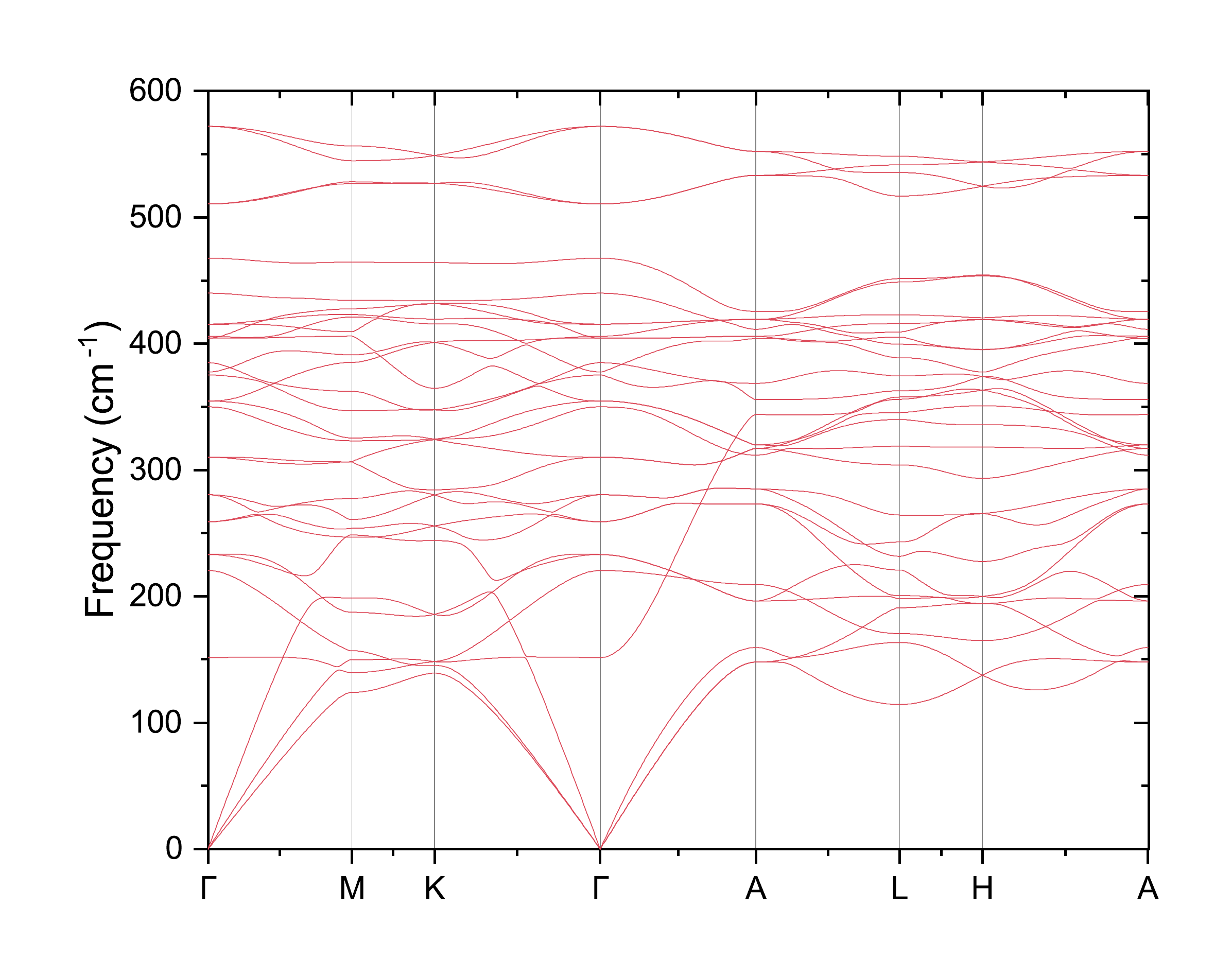}
	\centering
	\caption{The phonon spectrum of FePS$_3$ in space group $P\overline{3}1m$ at $V=250$ \AA$^3$, corresponding to pressure about $29$ GPa.}
	\label{fig:ph}
\end{figure}

Due to the lack of details of the short-range magnetic order and its breaking of translational invariance, it is beyond the scope of this work and further work is needed to reveal the short-range magnetic order at high pressure. Although our work on primitive $P\overline{3}1m$ shows nonmagnetic metallic state, which contradicts with the experimental finding, two things deserve to be mentioned here. One is the calculated phonon spectrum of $P\overline{3}1m$ in nonmagnetic state ($V=250$ \AA$^3$) does not show imaginary modes (see Fig.\ref{fig:ph}), indicating that the nonmagnetic metallic phase is stable and could exist under suitable conditions. Another is that we tried a supercell calculation of $P\overline{3}1m$ with $k=(010)$ (Fig.\ref{fig:structure}(c), with bottom two irons being `up' and top two irons `down'), which breaks $D_{3d}$ symmetry and is actually $C2/m(\beta=90^\circ,k=(010))$, at $V=260$ and $V=280$ \AA$^3$, respectively. For $V=260$ \AA$^3$, which is smaller than the discontinuity point, the crystal structure remains the same as that of $k=(000)$ and the magnetic moment vanishes. While for $V=280$ \AA$^3$, which is larger than the discontinuity point, the crystal structure recovers to $C2/m(\beta\sim90^\circ,k=(010))$ (see Fig.\ref{fig:EV}). It's possible that more complex magnetic structure in $P\overline{3}1m$ might further reduce the energy than $k=(010)$, but how to handle the magnetic symmetry in first-principle calculation would be challenging.

\section{Conclusions}
\label{sec:conclusion}
In conclusion, we have studied the magnetic structure transition of layered antiferromagnet FePS$_3$ under high pressure up to $50$ GPa. By preserving the magnetic point group of the crystal structures in the calculations, the ambiemt pressure structure is predicted to be $C2/m (\beta\sim107^\circ)$ with magnetic propagation vector $k=(01\frac{1}{2})$, which turns to $k=(010)$ at $0.7$ GPa ($U=2.5$ eV). The shift of neighboring layers is predicted to occur at $10.0$ GPa. These findings agree well with the experimental findings, except that we predict a metallic state with vanishing magnetic moment at $27.5$ GPa, while the experimental work reveals a metallic state with short-range magnetic order at $14$ GPa. The phase transition (3) and (4) also agree with previous theoretical works, though the predicted transition pressures are different, due to the different U value used in this work, which is found to have a profound influence on the predicted ground state and phase transitions. 

Besides, we report a new isostructural phase transition within $C2/m(\beta\sim107^\circ,k=(010))$ phase, caused by a rotation of Fe-S octahedron about $0.5^\circ$ through the $[001]$ axis. The accompanying  change of Brillouin zone and electronic properties during the phase transitions are analyzed. The Fe-S bond length is found to show a sudden decrease to $2.20$ \AA, accompanying the vanishing of magnetic moment. The exposition of the short-range magnetic order is beyond the scope of this work. Finally, note that we only considered the four related structures identified by experiment, and other low enthalpy structures can not be ruled out theoretically. Recent works have confirmed\cite{FePS3_PH1,Lee2016} that a lower energy state can be found by breaking the mirror symmetry of monolayer FePS$_3$, and a more detailed study on crystal searching is needed in the future. Our work highlights the importance of considering the detailed magnetic structure to arrive at reliable results in the research of layered antiferromagnet under high pressure.


\begin{acknowledgments}
The authors would like to acknowledge the computational support provided by National Supercomputer Center in Guangzhou and the Supercomputing Center of Hangzhou Dianzi University. Y. Z. thanks R. Xie and L. Wang for helpful discussions. This work is supported by Zhejiang Provincial Natural Science Foundation (LQ21A040010), the National Natural Science Foundation of China (NSFC-22090041,NSFC-11974432,NSFC-92165204 ),  NKRDPC-2018YFA0306001, Grants No. NKRDPC-2022YFA1402802,  GBABRF-2019A1515011337,
Leading Talent Program of Guangdong Special Projects (201626003), and International Quantum Academy of Shenzhen (Grant No. SIQSE202102).
\end{acknowledgments}

\bibliography{FePS3_bib}

\end{document}